\documentclass[aip, jcp, reprint]{revtex4-2}

\usepackage{graphicx}
\usepackage{amsmath}
\usepackage{physics}
\usepackage{bm}
\usepackage{amssymb}
\usepackage{multirow}
\usepackage{url}
\usepackage{hyperref}

\usepackage{longtable}   
\usepackage{booktabs}    
\usepackage{csvsimple}   
\usepackage{array}

\usepackage{xcolor}

\begin{document}


\title[Interaction-Induced Symmetry Breaking in Circular Quantum Dots]{Interaction-Induced Symmetry Breaking in Circular Quantum Dots} 



\author{Andres Perez Fadon}
\email{andres.perez-fadon19@imperial.ac.uk}
\thanks{\\The following article has been submitted to the Journal of Chemical Physics. After it is published, it will be found at \url{https://pubs.aip.org/aip/jcp}\\Copyright (2024) A. Perez Fadon, G. Cassella, H. Sutterud, W.M.C. Foulkes. This article is distributed under a Creative Commons Attribution-NonCommercial-NoDerivs 4.0 International (CC BY-NC-ND) License.}

\author{Gino Cassella}
\author{Halvard Sutterud}


\author{W. M. C. Foulkes}
\affiliation{Department of Physics, Imperial College London, South
Kensington Campus, London SW7 2AZ, United Kingdom}


\date{\today}

\begin{abstract}
This paper investigates interaction-induced symmetry breaking in circular quantum dots.
We explain that the anisotropic ``static Wigner molecule'' ground states frequently observed in simulations are created by interference effects that occur even in the non-interacting limit.
They have nothing in common with the interaction-driven crystallization of the uniform electron gas described by Wigner.
This leads us to define the term Wigner molecule more carefully, via a finite analog of the spontaneous symmetry breaking that arises in the homogeneous electron gas when the interactions are strong.
According to this definition, the charge density patterns characteristic of true interaction-induced Wigner molecules can only be seen if a small symmetry-breaking perturbation is applied to a strongly interacting quantum dot.
A simple argument based on separation of variables into center-of-mas and internal coordinates shows that the strength of the perturbation required to produce a finite effect on the density tends to zero in the limit as the strength of the interaction tends to infinity.
We confirm computationally that interaction-induced Wigner molecules satisfying these two criteria exist.
The neural-network variational Monte Carlo method used in our simulations proves more accurate than the coupled cluster and diffusion Monte Carlo methods employed in previous benchmark calculations of quantum dots.
\end{abstract}


\maketitle 

\section{Introduction}
\label{chap:intro}

Two-dimensional (2D) quantum dots have been a topic of research and technological interest for many years.
They provide a simple system in which to explore semiconductor heterostructures,\cite{jacak2013quantum, ashoori1996electrons} electron-electron interactions and quantum Hall phases.\cite{jacak2013quantum, kouwenhoven1998quantum, sivan1989aharonov}
Quantum dot lasers have found widespread use in optical communications, sensing and metrology.\cite{grillot2020quantumdot}
The properties of quantum dots are easily tunable with electromagnetic fields,\cite{jauregui1993wigner} suggesting that dots may be useful as elementary qubits for quantum computation.\cite{bester2004theory, kloeffel2013prospects, loss1998quantum, eriksson2004spin, culcer2012valley}
Computationally, quantum dots have been studied using a wide range of methods, including unrestricted Hartree-Fock,\cite{reusch2003unrestricted, pfannkuche1993comparison, de2007spin} exact diagonalization,\cite{reimann2000formation, merkt1991energy, mikhailov2002quantum, mikhailov2002two, mikhailov2002quantum, szafran2004anisotropic} density functional theory (DFT),\cite{koskinen1997hund, borgh2005correlation, jiang2003density, hirose1999spin, feng2017optical} variational and diffusion quantum Monte Carlo (VMC and DMC),\cite{harju2002wigner, kandemir2005variational, kim2021coupled,lohne2011ab} configuration interaction,\cite{rontani2006full, blundell2010precise, climente2006effect} path-integral Monte Carlo,\cite{egger1999crossover} and coupled cluster with single, double, and perturbative triple excitations (CCSD(T)).\cite{heidari2007electronic, salihbegovic2022coupled, lohne2011ab, waltersson2013performance}
In particular, a great deal of attention has been paid to the formation of Wigner molecules.\cite{jauregui1993wigner, reusch2001wigner, reusch2003unrestricted, thakur2022wigner, li2023wigner, mikhailov2002two, ghosal2007incipient, reimann2000formation, yannouleas2004unified}

This paper aims to dispel some of the confusion in the literature on Wigner molecules, provide a sharper definition of the concept of a Wigner molecule, and demonstrate that interaction-induced Wigner molecules satisfying the new definition exist.
In the process, we use the recently developed neural-network variational Monte Carlo (NNVMC) method to solve the many-electron Schr\"{o}dinger equation for a range of quantum dots containing up to 20 electrons, providing new benchmark results of higher accuracy than the CCSD(T) and DMC results currently available.

In 1934, Wigner predicted that a uniform electron gas undergoes an interaction-driven phase transition as the electron density is lowered at zero temperature.
Although the neutralizing positive background charge density is forced to remain uniform, the electrons localize into a state with long-range crystalline order.\cite{wigner1934interaction,wigner1938effects}
Two-dimensional Wigner crystals were recently imaged optically for the first time.\cite{smolenski2021signatures}
A short explanation of the Wigner transition appears at the beginning of Sec.~\ref{chap:wigner}.

Computational studies of small circular quantum dots often yield ground states with broken rotational symmetry and sometimes reveal blob-like maxima in the electron density.
The number of blobs only rarely matches the number of electrons, but it is nevertheless tempting to identify them as localized electrons and interpret the state as a molecular analog of a Wigner crystal.\cite{jauregui1993wigner, reusch2001wigner, reusch2003unrestricted, thakur2022wigner, li2023wigner, mikhailov2002two, ghosal2007incipient, reimann2000formation, yannouleas2004unified}
Section \ref{chap:wigner} explains that the rotational symmetry breaking is in fact caused by a simple interference effect\cite{pederiva2000diffusion, reusch2001wigner, ghosal2007incipient}
and can be observed equally well in non-interacting systems.
It has nothing to do with electron-electron interactions, correlations or the Wigner transition.
As far as we are able to ascertain, nobody has ever seen a true interaction-induced, symmetry-breaking Wigner molecule in a circular quantum dot.
Section \ref{chap:wigner} concludes by discussing what a true Wigner molecule would look like and how to identify one.

Although simulations of quantum dots often produce non-circular electron densities,
these are not universal.
Some dots have circularly symmetric electron densities but show interesting patterns in the electron-electron pair-correlation function.
Ground states of this type are sometimes called rotating Wigner molecules and are viewed as analogous to the floating Wigner crystals discussed in papers on the electron gas.\cite{bishop1982electron, lewin2019floating}
The nomenclature is more appropriate in this case, but the appearance of rotating Wigner molecules is often ascribed to spontaneous symmetry breaking at a phase transition.
Quantum dots cannot undergo phase transitions because they are finite, and the structure in the pair-correlation function is present even when the ground state possesses the full rotational symmetry of the Hamiltonian.
Rotating Wigner molecules are introduced in Sec.~\ref{chap:wigner} and considered in more detail in Sec.~\ref{chap:model}.

Section \ref{chap:model} begins by introducing the continuum model Hamiltonian we use to describe quantum dots and then considers the weakly interacting limit, demonstrating computationally that the ground-state spin and orbital angular momenta agree with Ghosal's\cite{ghosal2007incipient} modified version of Hund's rules.
These rules predict the $L\ne 0$ interference patterns often described as static Wigner molecules.

Another source of confusion concerns the nature of the sudden changes in ground state sometimes observed as the strength of the harmonic confining potential is changed at fixed electron number.
As discussed in Sec.~\ref{chap:model}, all sudden changes in the quantum dot ground state are level crossings and should not be described as phase transitions.
Furthermore, they have nothing to do with the formation of static or rotating Wigner molecules.
The symmetry breaking observed in the ground-state density of a static Wigner molecule is not caused by a level crossing.

Section~\ref{chap:ferminet} provides a brief review of the NNVMC method and describes the calculations carried out.
Section~\ref{chap:results} presents our benchmark results for quantum dots with up to 20 electrons and demonstrates the existence of true static Wigner molecules caused by electron-electron interactions.
Section~\ref{chap:conclusion} concludes.

\section{Wigner Crystals and Wigner Molecules}
\label{chap:wigner}

\subsection{The Wigner Transition}

We begin by reviewing the physics of the Wigner transition in an electron gas.
Consider a finite, 2D, homogeneous electron gas in a box of side $d$ and area $A=d^{2}$ under periodic boundary conditions.
The Schr\"odinger equation separates into center-of-mass and difference coordinates.
The center-of-mass part of the Hamiltonian is
\begin{align}
  -\frac{\hbar^2}{2mN}\nabla_{\bm{r}_c}^{2},
\end{align}
where $N$ is the number of electrons, $m$ is the electron mass, and the Laplacian is taken with respect to the position vector $\bm{r}_c$ of the center of mass.
The normalized center-of-mass eigenfunctions are plane waves, $e^{i\bm{k}\cdot\bm{r}_c}/d$, and the center-of-mass ground state is at zero energy with $\bm{k}=\bm{0}$.
It follows that the ground state is invariant under simultaneous translations of all of the electrons.
The lowest-energy eigenstate with center-of-mass momentum $\hbar\bm{k}$ has energy $\hbar^2 k^2 / 2mN = \hbar^2 k^2 / 2mnA$, where $n$ is the electron number density.

At low enough density, the Coulomb potential dominates and long-range order develops in the wave function of the difference coordinates.
The new state is called a floating Wigner crystal\cite{bishop1982electron, lewin2019floating}  and is visible in the pair-correlation function.
Defining $\rho(\bm{r}_{1}, \bm{r}_{2})\equiv \frac{A}{N(N-1)}\widetilde{\rho}(\bm{r}_{1} - \bm{r}_{2})$, where $\rho$ is the pair-density function and $\widetilde{\rho}$ is normalized to unity, the Fourier component
\begin{align}
    F_{\bm{G}} &= \int d^{2}\bm{r} \widetilde{\rho}(\bm{r})e^{i\bm{r}\cdot\bm{G}},
\end{align}
where $G$ is a reciprocal lattice vector of the emergent floating lattice, remains finite even in the limit as $d\rightarrow\infty$ at constant $n=N/A$.
Hence, $F_{\bm{G}}$ acts as an order parameter for the phase transition.

Although the ground-state one-electron density is still uniform, the differences between the ground state energy and the energies of the low-energy excited states created by adding center-of-mass momentum tend to zero as $d\rightarrow\infty$ at constant electron density $n$.
Hence, in the thermodynamic limit, an arbitrarily small perturbation is sufficient to mix states with different center-of-mass $\bm{k}$ and localize the center of mass.
This interaction-induced spontaneous symmetry breaking produces a static Wigner crystal, visible in both the one-electron density and the pair-correlation function.

\subsection{Static Wigner Molecules as Interference Effects}

We now move on to investigate the formation of Wigner molecules in quantum dots.
Start with a single electron in a 2D parabolic potential and consider an energy eigenstate with $l \ne 0$, where $\hbar l$, $l \in \mathbb{Z}$, is the orbital angular momentum.
Note that $l$ may be negative in 2D, and that there is no 2D analog of the magnetic quantum number $m$.
Due to time-reversal symmetry, eigenstates with angular momentum quantum numbers $\pm l$ are degenerate.
Hence, any linear superposition of the states $\ket{+l}$ and $\ket{-l}$ is also an energy eigenstate.
The electron number density depends on the particular superposition chosen: if $\ket{\psi}=\alpha\ket{+l}+\beta\ket{-l}$, the angular density 
\begin{align}
    n(\theta) &= \frac{1}{2\pi}\left[1+2|\alpha\beta^{*}|\cos(2l\theta + \arg(\alpha\beta^{*}))\right]
\end{align}
has $2l$ angular maxima and $2l$ angular minima.

The ground state of a one-electron circular quantum dot has $l=0$ and is thus rotationally uniform, but $N$-electron quantum dots are analogous to atoms, where shell-filling effects often yield ground states with finite values of the total angular momentum.
The expectation value $L$ of the total angular momentum operator $\hat{L} = \sum_{i=1}^{N} \hat{l}_i$, where $\hat{l}_i$ is the angular momentum of electron $i$, is then non-zero.
As shown in Appendix \ref{chap:so2symm}, the one-electron density of any simultaneous eigenstate of the Hamiltonian $\hat{H}$ and the total angular momentum operator $\hat{L}$ must be circularly symmetric.
However, as in the one-electron case, a general linear combination of the degenerate eigenstates $\ket{\pm L}$ has a non-uniform angular density with $2L$ angular maxima and $2L$ angular mimina.
In particular, restricting to real wavefunctions will always produce a non-uniform angular density if the angular momentum is non-zero.
All simulations of circular quantum dots that have produced non-uniform angular densities have had ground states of this type.

\begin{figure}
    \centering
    \includegraphics[scale = 0.525]{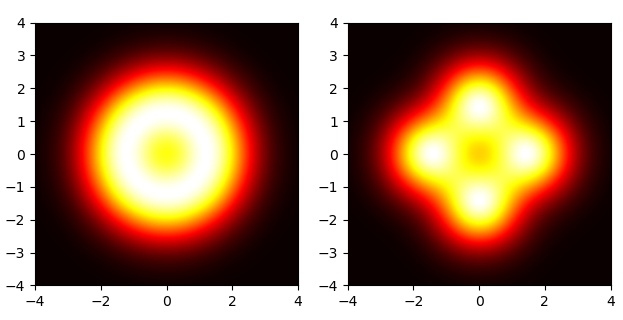}
    \caption{Electron densities for two degenerate ground states of a four-electron circular quantum dot in the non-interacting limit.
      Both states have total spin $S=2$.
      The system has time-reversal symmetry, so many-electron states with in-plane total orbital angular momentum $\pm L$ are degenerate.
      Left: circularly symmetric electron density, corresponding to an angular momentum eigenstate with $L = +2$.
      Right: electron density with angular modulation, corresponding to an equal superposition of states with angular momentum $L = +2$ and $L = -2$.
      Distances are measured in units of $l_{0}$, defined in Eq.~(\ref{eq:l0}).
    }
    \label{fig:pdf_real_choice}
\end{figure}

Figure~\ref{fig:pdf_real_choice} shows two possible ground-state electron densities for a non-interacting four-electron quantum dot with total spin $S=2$: the ground state without rotational symmetry in the single-particle density, shown on the right, is constructed by superposing degenerate ground states with equal and opposite total angular momentum $L = \pm 2$.
Quantum dot ground states of this type are often called (static) Wigner molecules.\cite{yannouleas1999spontaneous, reimann2002electronic, mitroy2013theory, yannouleas2004unified, sheikh2021symmetry, rontani2006full}
However, as pointed out in Refs.~\onlinecite{pederiva2000diffusion, reusch2001wigner, ghosal2007incipient}, the non-uniformity is caused by the $\pm L$ interference effect and has nothing to do with interactions or correlations.

\subsection{True Wigner Molecules}

If the confining potential binding the electrons to a quantum dot is weak, the electrons are far apart on average and the interaction energy, although small, is larger than the kinetic energy.
This is the strongly interacting regime.
If the confining potential is strong, the electrons are close together, the kinetic energy dominates, and the dot is weakly interacting.
One might hope to encourage the formation of true Wigner molecules by decreasing the confining potential and thus increasing the effective strength of the Coulomb interaction of the quantum dot.

Unfortunately, a simple symmetry argument (see Appendix \ref{chap:so2symm}) shows that the single-particle density of any simultaneous eigenstate of the Hamiltonian $\hat{H}$ and the total orbital angular momentum $\hat{L}$ must remain circularly symmetric regardless of the strength of the interaction.
Weakening the confining potential is not sufficient to create static Wigner molecules. 
Instead, as the interaction strengthens, angular structure develops smoothly in the pair-correlation function.\cite{ghosal2007incipient}
Ground states with molecule-like structure in the pair-correlation function but not the one-electron density are sometimes called rotating Wigner molecules.

In the limit as the strength of the interaction tends to infinity, Appendix \ref{chap:separation} shows that the separation in energy between states with different values of the center-of-mass angular momentum $l_c$ tends to zero. (Note that $l_c$ is not the same as the total angular momentum $L$.)
Therefore, any small $SO(2)$-breaking perturbation mixes states with different center-of-mass angular momenta.
The rotating Wigner molecule becomes locked into its lowest energy orientation and the angular structure, previously seen only in the pair-correlation function, becomes visible in the one-electron density.
The stronger the interaction, the smaller the energy difference between states with different center-of-mass orbital angular momenta, the weaker the perturbation required to localize the rotating Wigner molecule.

The symmetry-broken state created by the small perturbation is a true static Wigner molecule. Its electron density is localized into $N$ blobs ($N$ being the number of electrons), independent of the total angular momentum $L$.
Figure~\ref{fig:perturbation} in Sec.~\ref{chap:results} shows the characteristic electron density of a true static Wigner molecule in two examples of quantum dots with $L=0$ ground states.
Rotating and true static Wigner molecules are induced by strong Coulomb interactions and are not trivial interference effects. 

Unfortunately, applying a small symmetry-breaking perturbation also picks out a specific linear combination of the degenerate eigenstates $\ket{+L}$ and $\ket{-L}$, polluting the density signature of the Wigner crystal with that of the interference effect.
This complicates the electron densities observed in strongly interacting quantum dots with $L \neq 0$.

\section{Static and Rotating Wigner Molecules in a Model Quantum Dot}
\label{chap:model}

\subsection{Hamiltonian}

We model quantum dots by the Hamiltonian
\begin{align}
  \hat{H} &= \sum_{i=1}^{N}\left(-\frac{\hbar^{2}}{2m}\nabla_{i}^{2} + \frac{1}{2}m\omega^{2}r_{i}^{2}\right) + \sum_{i=1}^{N}\sum_{j>i}^{N}\frac{e^{2}}{4\pi\epsilon_{0}|\bm{r}_{i} - \bm{r}_{j}|},\label{eq:H_og}
\end{align}
where $m$ is the electron mass, $N$ is the number of electrons, $\omega$ is the harmonic potential strength, $e$ is the electron charge, $\epsilon_{0}$ is permittivity of free space, and $\bm{r}_{i}$ is the position of electron $i$.
This Hamiltonian can be made dimensionless by defining
\begin{align}
  l_{0} &= \sqrt{\frac{\hbar}{m\omega}}\label{eq:l0}, & 
    \lambda &= \frac{e^{2}}{4\pi\epsilon_{0}l_{0}\hbar\omega},
\end{align}
where $l_{0}$ has dimensions of length and $\lambda$ is dimensionless.
Expressing distances in units of $l_{0}$ and dividing by $\hbar\omega$, we obtain
\begin{align}
  \hat{\mathcal{H}} &= \frac{\hat{H}}{\hbar\omega} = \sum_{i=1}^{N}\left(-\frac{1}{2}\nabla_{i}^{2}+\frac{1}{2}r_{i}^{2}\right) + \sum_{i=1}^{N}\sum_{j>i}^{N}\frac{\lambda}{|\bm{r}_{i} - \bm{r}_{j}|}\label{eq:H},
\end{align}
where $\lambda$ is interpreted as an effective Coulomb interaction strength, and we have introduced $\bm{r}^{\prime}=\bm{r}/l_{0}$ and then relabeled $\bm{r}^{\prime}\rightarrow\bm{r}$.
All variables in Eq.~(\ref{eq:H}) are dimensionless.
This is the form of the Hamiltonian used for our NNVMC calculations.

\subsection{Hund's Rules}

The one-electron energy eigenfunctions of a 2D harmonic oscillator with $m = \omega = 1$ are known as Darwin-Fock states and take the form\cite{kouwenhoven2001few}
\begin{align}
    \psi_{nl}(r, \theta) &= \frac{e^{il\theta}}{\sqrt{\pi}}\sqrt{\frac{n!}{(n+|l|)!}}e^{-r^{2}/2}r^{|l|}L_{n}^{|l|}(r^{2}),
\end{align}
where $n\in\mathbb{N}$ is the principal quantum number, $l\in \mathbb{Z}$ is the 2D orbital angular momentum, and $L_{n}^{|l|}$ is a generalized Laguerre polynomial.
The corresponding energy eigenvalues are 
\begin{align}
    E_{nl} &= 2n + |l| + 1.
    \label{eq:energy-levels}
\end{align}
Unlike in 3D, the value of $|l|$ may exceed $n$.

In the limit as $\lambda \rightarrow 0$, the $N$-electron ground state of the Hamiltonian from Eq.~(\ref{eq:H}) is a Slater determinant of the $N$ single-particle orbitals of the lowest energy.
If the highest occupied one-electron orbital is degenerate and $\lambda$ is finite but small, our simulations confirm that $N$-electron quantum dots satisfy the modified version of Hund's rules proposed by Ghosal \emph{et al.}:\cite{ghosal2007incipient} the ground state is the one that maximizes the total spin, and then, if there is a tie, orbitals are occupied in descending order of  $|l|$, thus maximizing $\sum_{i\;\text{occ}} |l_i|$.
This differs from the 2D analog of the rule for atoms, which would maximize $|L| = |\sum_{i\;\text{occ}} l_i|$.

\begin{figure}
    \centering
    \includegraphics[scale = 0.14]{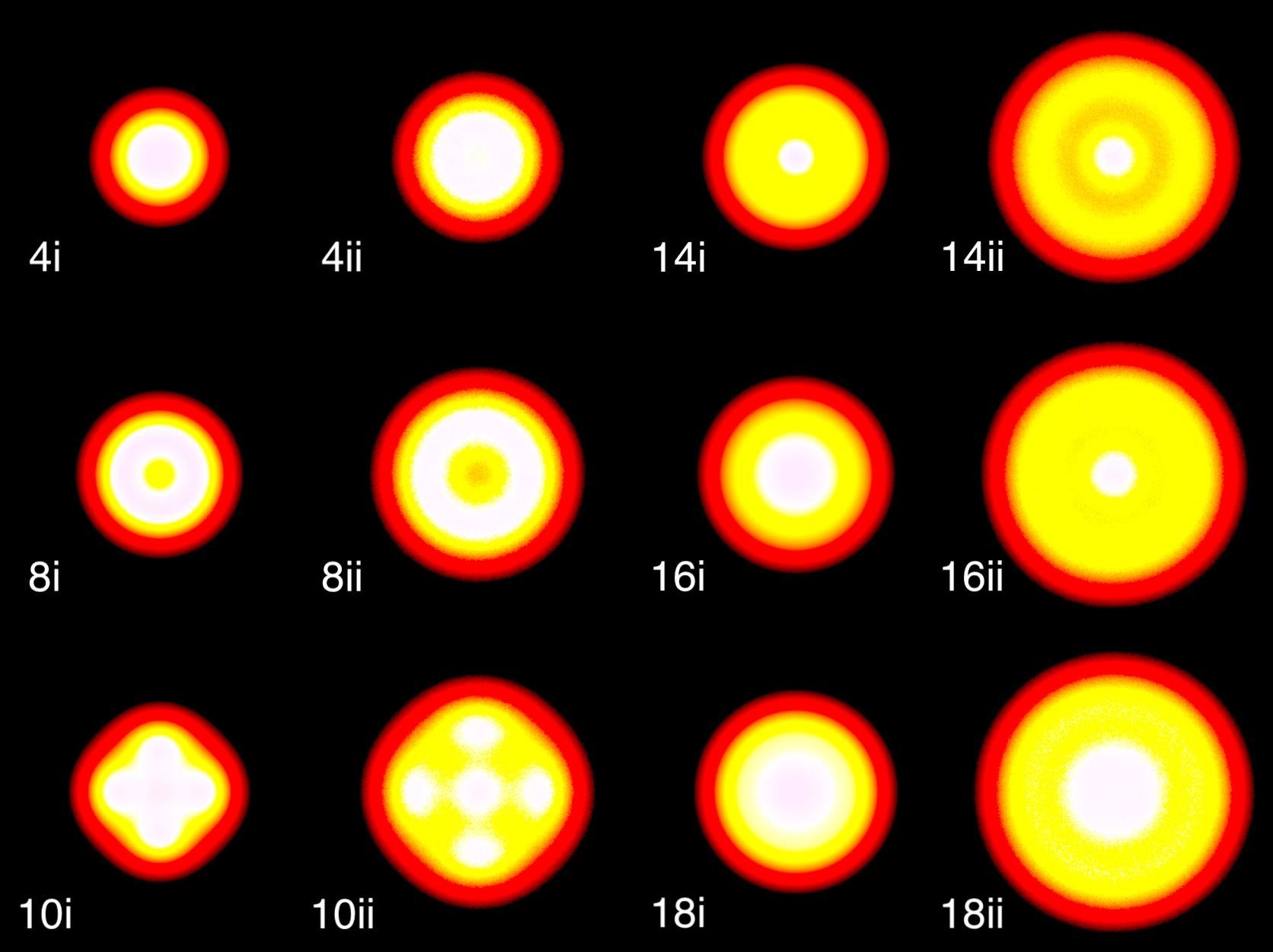}
    \caption{Electron densities for unpolarized quantum dots.
      The subfigure labels indicate the number of electrons in the quantum dot.
      Densities labeled `i' were calculated from a Slater determinant of non-interacting one-electron orbitals occupied according to the modified Hund's rules.
      Densities labeled `ii' were computed using the NNVMC method at $\lambda = 1$.
      All subfigures show the region $-4l_{0} \leq x, y \leq 4l_{0}$ and all densities are obtained from real superpositions within every degenerate subspace.
    }
    \label{fig:qdots}
\end{figure}

Consider as an example the 10-electron, spin unpolarized quantum dot.
We use the following notation: $(n, l)_{\sigma}$ indicates the principal quantum number $n$, the 2D orbital angular momentum $l$, and the spin projection $\sigma$ of a one-electron orbital; whilst $\{L, S\}$ indicates the total angular momentum and total spin of an $N$-electron energy eigenstate. In the limit $\lambda\rightarrow 0$, the ground state is a Slater determinant of the following single-particle orbitals: $(0, 0)_{\uparrow}$, $(0, 0)_{\downarrow}$, $(0, 1)_{\uparrow}$, $(0, -1)_{\uparrow}$, $(0, 1)_{\downarrow}$, $(0, -1)_{\downarrow}$, $(0, 2)_{\uparrow}$, $(0, -2)_{\uparrow}$, $(1, 0)_{\uparrow}$, $(0, 2)_{\downarrow}$, which has $\{L, S\} = \{2, 1\}$. Therefore, if one works with real superpositions, four maxima should be seen in the single particle density, as seen in Fig.~\ref{fig:qdots}.

Figure~\ref{fig:qdots} compares one-electron densities for unpolarized quantum dots obtained: (i) by occupying non-interacting one-electron orbitals according to the  modified Hund's rules; and (ii) by using NNVMC simulations to compute the interacting ground state at $\lambda=1$.
Real wave functions are used in both cases; if complex eigenfunctions of $\hat{L}$ had been used, all densities would have been circularly symmetric.
The interactions push the electrons in the NNVMC simulations further apart, but the qualitative agreement between the non-interacting densities computed using the modified Hund's rules and the interacting densities computed using NNVMC is excellent.

\subsection{Level Crossings}

A simple example of a level crossing is seen the $N=3$ quantum dot.
The non-interacting $S=3/2$ ground state is obtained by occupying the $(0, 0)_{\uparrow}$, $(0, 1)_{\uparrow}$ and $(0, -1)_{\uparrow}$ orbitals, for a total non-interacting energy of 5$\hbar\omega$.
The non-interacting $S=1/2$ ground state is obtained by occupying the $(0, 0)_{\uparrow}$, $(0,0)_{\downarrow}$ and $(0, 1)_{\uparrow}$ orbitals, for a total non-interacting energy of 4$\hbar\omega$.
Hence, for small enough values of $\lambda$, the overall ground state has quantum numbers $\{L, S\} = \{1, 1/2\}$.

Hund's first rule indicates that, in the presence of Coulomb interactions, the total spin will be maximized.
A rewording of this statement is that the Coulomb energy, when treated as a perturbation, is lower the larger the total spin eigenvalue.
Therefore, as we increase $\lambda$, the perturbation overtakes the difference in energy between the $\{1, 1/2\}$ state and the $\{0, 3/2\}$ state.
The two states cross and the $L$ and $S$ quantum numbers of the ground state change discontinuously at some specific interaction strength $\lambda_c$.
An exact-diagonalisation study showing this crossing as a function of $\lambda$ can be found in Ref.~\onlinecite{mikhailov2002quantum}.

As far as we have been able to ascertain, all of the quantum dot ``phase transitions'' discussed in the literature~\cite{yannouleas2004unified, yannouleas2007symmetry, muller1996phase} are level crossings of this type. 
%
%
Furthermore, although changes in $L$ resulting from level crossings may produce ground states with densities that depend on angle, suggesting a Wigner transition, the non-uniformity is caused by the $\pm L$ interference effect and disappears when complex eigenfunctions of $\hat{H}$ and $\hat{L}$ are used.
It never indicates the appearance of a true Wigner crystal.

The perturbative argument above explains why level crossings may happen as $\lambda$ increases, but the argument is only reliable when the perturbing potential is small in comparison with the eigenvalue separation, in which case it cannot make up the energy difference between levels.
We are pushing the argument beyond its range of validity and transitions may or may not happen in practice.
Indeed, in the limit as $\lambda \rightarrow \infty$, we expect levels with different values of $S$ to become degenerate.
This is because the Hamiltonian is spin independent, so the spin only matters in the form of fermionic exchange.
As the strength of the Coulomb interaction diverges, the electrons are driven arbitrarily far apart, their wave functions no longer overlap, and their relative spins no longer affect the energy.\cite{reusch2003unrestricted}

\subsection{Wigner Molecules Revisited}
The literature on rotating Wigner molecules contains inconsistencies, similar to the case of static Wigner molecules.
They are often said to exist only after a transition, in the regime where $\lambda > \lambda_{c}$,\cite{yannouleas2004unified, mikhailov2002two, egger1999crossover, reusch2001wigner, mikhailov2002quantum} and the state with $\lambda > \lambda_c$ is often called a symmetry-broken phase.\cite{thakur2022wigner, yannouleas2004unified, egger1999crossover}
The implication is that the structure in the pair-correlation function is caused by a symmetry-breaking phase transition.
We do not consider this to be an appropriate definition for the following reasons:
\begin{enumerate}
    \item If any properties of the ground state change discontinuously at some interaction strength $\lambda_c$, it must be due to a level crossing.
      Phase transitions can only happen in the thermodynamic limit.\cite{beekman2019introduction}
    \item In some quantum dots, the ground state has the same total spin and angular momentum eigenvalues for all studied values of $\lambda$,\cite{mikhailov2002two} and hence could never become a rotating Wigner molecule under this definition.
      However, the properties of such dots at high $\lambda$ are indistinguishable from those of rotating Wigner molecules.
    \item All quantum dots with $N > 1$ show angular structure in the pair-correlation function. The structure strengthens as $\lambda$ increases and may change discontinuously at level crossings, but does not appear suddenly.
    \item Level crossings can happen at relatively small $\lambda$, where the Coulomb interaction is not yet dominant.
    \item The state that becomes the ground state for $\lambda>\lambda_{c}$ still exists when $\lambda < \lambda_{c}$, although it is not in general the ground state in that regime.
\end{enumerate}

A better way to think about rotating and static Wigner molecules uses the idea of spontaneous symmetry breaking adapted for finite systems.
Let us first recap the essential features of the Wigner transition of a uniform electron gas, as discussed in Sec.~\ref{chap:wigner}.
The ground state is invariant under simultaneous translation of all of the electron positions (\emph{i.e.}, the center of mass has $\bm{k}=\bm{0}$), but the energy required to boost the center-of-mass momentum from $\bm{0}$ to $\bm{k}$ scales as $N^{-1} \propto A^{-1}$.
In the thermodynamic ($N \rightarrow \infty$) limit, the ground state becomes degenerate with states of any $\bm{k}$ and an arbitrarily small local perturbation is sufficient to collapse the state into a Wigner crystal, breaking the translational symmetry of the Hamiltonian.

A closely analogous way to identify a true Wigner molecule is to check whether the following four criteria are obeyed:
\begin{enumerate}
    \item Except at level crossings, which are not directly relevant to Wigner molecule formation, the properties of quantum dots are continuous functions of the strength of the Coulomb interaction.
      The features identified with Wigner molecules appear smoothly as $\lambda$ increases ($\omega$ decreases).
    \item If no symmetry-breaking perturbation is present and the ground state is chosen to be a simultaneous eigenstate of total energy and total orbital angular momentum, the single-particle density is rotationally invariant.
      The pair-correlation function still has interesting stuctures, which strengthen gradually as $\lambda$ increases.\cite{ghosal2007incipient}
      If the modulation of the pair-correlation function is strong, we call such systems rotating Wigner molecules.
    \item When the interaction strength $\lambda$ is large, energy eigenstates with many different center-of-mass angular momenta lie very close in energy to the ground state.
      (This is shown mathematically in Appendix~\ref{chap:separation}.)
      A small symmetry-breaking perturbation can mix these states, allowing the construction of highly angularly localized states we call static Wigner molecules.
      The angular structure already present in the pair-correlation function then becomes visible in the one-electron density.
      The observed structure depends on the number of electrons occupying the quantum dot, not its total angular momentum.
    \item In the limit as $\lambda \rightarrow \infty$, the symmetry-breaking perturbation required to create a static Wigner molecule becomes arbitrarily small.
\end{enumerate}
Explicit computational examples of condition 3 are presented in the next section.

If one replaces `Wigner molecule(s)' by `Wigner crystal(s)', `quantum dots' by `electron gas simulation cells', `orbital angular momentum' by `linear momentum', `rotationally' by `translationally',  `rotating' by `floating', `center-of-mass angular momenta' by `center-of-mass momenta', `angularly' by `translationally', and `angular' by `translational', these criteria also describe the Wigner crystallization of a finite simulation cell of uniform electron gas subject to periodic boundary conditions.
In that case, however, instead of letting the strength of the Coulomb interaction tend to infinity, we can let the number of electrons in the system tend to infinity, increasing the size of the periodic box as required to keep the electron density fixed.
If the electron density lies below some critical value, the ground state of the infinite system remains crystalline no matter how weak the symmetry-breaking perturbation; if the electron density lies above the critical value, the degree of crystallinity (the order parameter) tends to zero as the strength of the perturbation tends to zero.
Unlike Wigner molecule formation, Wigner crystallization in an infinite system is a true phase transition.

\section{The Neural-Network Variational Monte Carlo Method}
\label{chap:ferminet}

The difficulty of computing the properties of interacting many-particle quantum systems lies in the exponential growth of the size of the Hilbert space with particle number.
Solutions for systems of a few particles moving in a few specific potentials are known exactly, but computational methods are required to tackle more complex problems.

One such method, variational Monte Carlo (VMC), relies on the use of a parameterized trial ground-state wave function.
The expectation value of the Hamiltonian depends on the wave function parameters, the optimal values of which may be found by minimizing the expected energy using gradient descent or other optimizers.
If the ansatz is sufficiently flexible, the minimal energy expectation value is a good upper bound on the true ground-state energy and the corresponding trial wave function approximates the true ground state.

Traditional implementations of VMC require the user to choose a finite basis set of one-electron functions.
Linear superpositions of the basis functions define single-particle orbitals, which are used to construct Slater determinants.
The ansatz is a linear combination of Slater determinants multiplied by a Jastrow factor, and the optimization determines the best available orbitals, determinant coefficients, and Jastrow parameters.

In practice, the choice of finite basis set often limits the quality of the trial wave function.
An alternative approach that has had great success in recent years replaces the orbitals in the Slater determinants by deep neural networks, which can be evaluated without choosing a specific basis set.
The FermiNet\cite{ferminet_github} network architecture used here has had remarkable success predicting the ground and excited states of molecules and solids,\cite{pfau2020ab, spencer2020better, pfau2024natural, lu2023penalty, li2022ab} phase transitions,\cite{cassella2023discovering} positronic chemistry,\cite{cassella2024neural} and superfluids.\cite{lou2024neural}
Calculations can also be accelerated using pseudopotentials\cite{li2022fermionic} and forward Laplacian evaluation.\cite{li2024computational}
Alternative network architectures include PauliNet,\cite{hermann2020deep, entwistle2023electronic} DeepQMC,\cite{schatzle2023deepqmc} NetKet\cite{vicentini2022netket} (mostly for lattice models) and modifications,\cite{choo2020fermionic} the Psiformer self-attention approach,\cite{von2022self} and more.\cite{scherbela2022solving, scherbela2024towards, wilson2021simulations}

A distinguishing feature of most NNVMC methods is that the orbitals no longer depend on the coordinates of just one electron.
In addition to the dependence on a particular electronic position $\bm{r}_{j}$ and spin $\sigma_{j}$, they also depend on the positions and spins of all of the other electrons:
\begin{align}
    \Psi &= \sum_{k}^{n_{\text{det}}}a_{k}\det_{ij}[\psi^{k\uparrow}_{i}(j)]\det_{ij}[\psi^{k\downarrow}_{i}(j)],
    \label{eq:ferminet-wavefunction}
\end{align}
where $i$ and $j$ label the rows and columns of the Slater matrices and $k$ indexes the Slater determinants evaluated from those matrices.
Here
\begin{align}
    \psi_{i}^{k\sigma}(j) &\equiv \psi_{i}^{k\sigma}(\bm{r}^{\sigma}_{j}, \{\bm{r}^{\sigma}_{/j}\}, \{\bm{r}^{\overline{\sigma}}\}),
    \label{eq:ferminet-orbital}
\end{align}
where $\sigma \in \{\uparrow, \downarrow\}$ is a spin label, $\{\bm{r}^{\sigma}_{/j}\}$ denotes the set of positions of all spin-$\sigma$ electrons excluding electron $j$, and $\{\bm{r}^{\overline{\sigma}}\}$ is the set of positions of all electrons of the opposite spin.
As the use of set notation suggests, the value of the orbital in Eq.~(\ref{eq:ferminet-orbital}) is invariant under interchanges of the elements in $\{\bm{r}^{\sigma}_{/j}\}$ or $\{\bm{r}^{\overline{\sigma}}\}$.
Any wave function in the form of Eq.~(\ref{eq:ferminet-wavefunction}) obeys Fermi-Dirac exchange statistics, while the general nature of the many-electron orbitals improves the expressivity of the ansatz and reduces the number of determinants required in the linear combination.\cite{pfau2020ab}

The orbitals $\psi_{i}^{k\sigma}$ are represented as a function, $\phi^{k\sigma}_{i}$, generated by a neural network, modulated by an envelope, $f^{k\sigma}_{i}$:
\begin{align}
    \psi^{k\sigma}_{i}(j) &= f^{k\sigma}_{i}(j)\phi^{k\sigma}_{i}(j),
\end{align}
where 
\begin{align}
    f_{i}^{k\sigma}(j) &= \exp[-(\Xi^{k\sigma}_{i} r^{\sigma}_{j})^{2}],
\end{align}
and the $\Xi{i}^{k\sigma}$ are learnable parameters.
The neural network takes the electron positions and spins as inputs and returns the corresponding values (and derivatives) of all of the $\phi_i^{k\sigma}(j)$.
Gaussian envelopes were used as the single-electron eigenstates of a harmonic circular quantum dot include a Gaussian factor.
This allows the envelope to capture the behavior of the wave function far from the center of the harmonic well, where the parabolic potential dominates over the Coulomb interaction.
The structure of the deep neural network that generates the $\phi_{i}^{k\sigma}(j)$ is described in Refs.~\onlinecite{pfau2020ab, spencer2020better}.

When we wish to explore condition 3 in the definition of a Wigner molecule, we add the following rotational symmetry breaking perturbation to the Hamiltonian $\hat{H}$ from Eq.~(\ref{eq:H_og}),
\begin{align}
  \hat{P} &= -\mu\sum_{i}\frac{1}{\sqrt{2\pi}\sigma}\exp\left[\frac{-|\bm{r}_{i} - \bm{r}_{0}|^{2}}{2\sigma^{2}}\right],
\end{align}
where $\bm{r}_{0}$ is chosen to lie at the first maximum away from the origin of the radial density $n(r)$.
Dividing by $\hbar\omega$ and measuring distances in units of $l_{0}$, the perturbation added to $\hat{\mathcal{H}}$ in Eq.~\ref{eq:H} becomes
\begin{align}
  \hat{\mathcal{P}} &= -\eta\lambda\sum_{i}\frac{1}{\sqrt{2\pi}\sigma_{0}}\exp\left[\frac{-|\bm{r}_{i}-\bm{r}_{0}|^{2}}{2\sigma_{0}^{2}}\right]\label{eq:perturbation},
\end{align}
where $\eta=\mu 4\pi\epsilon_{0}/e^{2}$ and $\sigma=\sigma_{0}l_{0}$.
Both $\eta$ and $\sigma_{0}$ are dimensionless.
Note that $\hat{\mathcal{P}}$ grows linearly with $\lambda$. Appendix~\ref{chap:separation} shows that the separation of energy levels with different center-of-mass angular momentum is of order $\hbar\omega$.
In the dimensionless form used here, where energies are expressed in units of $\hbar\omega$, this becomes a separation of order unity for all $\lambda\ge 0$.
Hence, no matter how small the strengthen $\mu$ of the SO(2)-breaking perturbation, it will be sufficient to mix states of different center-of-mass angular momentum and freeze the rotating Wigner molecule when the interaction strength $\lambda$ is large enough. 
This is in agreement with condition 3.

\section{Computational Results}
\label{chap:results}

We are unaware of any previous work on circular quantum dots using NNVMC, so we compare our results against benchmarks computed using two other highly accurate methods: DMC and CCSD(T).\cite{lohne2011ab}
The DMC and NNVMC methods are variational, so the computed energy is an upper bound on the exact ground-state energy and calculations that yield lower energies are better.
CCSD(T) is not variational and may yield results below the exact ground-state energy.
Nevertheless, if the NNVMC result happens to be lower than the CCSD(T) result, the NNVMC result is closer to the exact energy.
Surprisingly, although the CCSD(T) energies from Ref.~\onlinecite{lohne2011ab} were not extrapolated to the complete-basis-set (CBS) limit, they lie below the CCSD(T) energies from Ref.~\onlinecite{salihbegovic2022coupled}, which were extrapolated.
We therefore compare with Ref.~\onlinecite{lohne2011ab}.
As shown in Table \ref{table:energies}, the ground-state energies obtained with NNVMC are consistently better than the DMC and CCSD(T) energies.

\begin{table}
\centering
\begin{tabular}{|c c|c c c|}
  \hline
  $N$ & $\omega$ & \text{FermiNet} & \text{CCSD(T)} & \text{DMC}\\
  \hline
  \hline
  \multirow{2}{*}{2} & 0.5 & \textbf{1.659 768(5)} & \textbf{1.659 772} & \textbf{1.659 75(2)}\\
  & 1.0 & \textbf{3.000 00(2)} & \textbf{3.000 000} & \textbf{3.000 00(3)}\\
  \hline
  \multirow{3}{*}{6} & 0.28 & \textbf{7.597 44(5)} & 7.600 6 & 7.600 1(1)\\
  & 0.5 & \textbf{11.781 3(1)} & 11.783 7 & 11.788 8(2)\\
  & 1.0 & \textbf{20.154 8(1)} & 20.157 0 & 20.159 7(2)\\
  \hline
  \multirow{3}{*}{12} & 0.28 & \textbf{25.626 34(8)} & 25.632 4 & 25.635 6(1)\\
  & 0.5 & \textbf{39.146 7(1)} & 39.151 6 & 39.159(1)\\
  & 1.0 & \textbf{65.684 6(2)} & 65.688 6 & 65.700(1)\\
  \hline
  \multirow{3}{*}{20} & 0.28 & \textbf{61.910 3(2)} & 61.915 6 &  61.922(2)\\
  & 0.5 & \textbf{93.852 6(3)} & 93.855 8 & 93.867(3)\\
  & 1.0 & \textbf{155.849 2(4)} & 155.857 1 & 155.868(6)\\
  \hline
\end{tabular}
\caption{NNVMC energies of unpolarized ($S_{z}=0$) quantum dots obtained with the FermiNet ansatz.
  Also shown are DMC results and CCSD(T) results computed in a basis set including 20 major oscillator shells. 
  (There are no triples when $N=2$, so CCSD(T) reduces to CCSD in that case.)
  All energies are expressed in Hartrees.
  Bold values represent the best result for a given electron number and harmonic potential strength $\omega$.
  The CCSD(T) and DMC values are taken from Ref.~\onlinecite{lohne2011ab}.\label{table:energies}
}
\end{table}

\begin{figure}
    \centering
    \includegraphics[scale = 0.55]{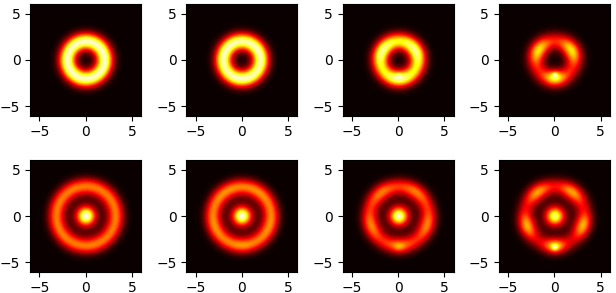}
    \caption{Effect of the $SO(2)$-breaking perturbation defined in Eq.~(\ref{eq:perturbation}).
    Top row: $N=3$, $S_{z}=3/2$, $L=0$, $\lambda = 15$.
  Bottom row: $N=6$, $S_{z}=0$, $L=0$, $\lambda=15$.
Columns represent different values of $\eta$, from left to right: 0, 0.01, 0.05, 0.1.
Lengths are measured in units of $l_{0}$.}
    \label{fig:perturbation}
\end{figure}

We now consider adding the symmetry-breaking perturbation defined in Eq.~(\ref{eq:perturbation}).
All calculations were made with $\sigma_{0}=0.1$.
The results reported in Table \ref{table:eta-min} show that, as $\lambda$ increases, smaller values of $\eta$ become sufficient to produce a ground state with substantial symmetry breaking.
We measure the symmetry breaking by first calculating the angular single-particle density $n(\theta)$, measuring its root mean square deviation (RMSD) over one period, and comparing against a critical value, chosen to be 0.01.
\begin{table}
\begin{center}
\begin{tabular}{|c c c|c|}
  \hline
  \quad $N$ \quad & \quad $S_{z}$ \quad & \quad $\lambda$ \quad & \quad $\eta_{\text{min}}$ \quad\\
  \hline
  \hline
  \multirow{5}{*}{3} & \multirow{5}{*}{3/2} & 1 & $>$1\\
  & & 5 & 0.5\\
  & & 10 & 0.1\\
  & & 15 & 0.1\\
  & & 20 & 0.05\\
  \hline
  \multirow{5}{*}{6} & \multirow{5}{*}{0} & 1 & $>$1\\
  & & 5 & 0.5\\
  & & 10 & 0.1\\
  & & 15 & 0.05\\
  & & 20 & 0.05\\
  \hline
\end{tabular}
\caption{Minimum value of $\eta$ necessary to collapse different quantum dots into a symmetry-broken ground state.
We refer to a state as ``symmetry-broken" if the root mean square deviation (RMSD) of the angular single-particle density $n(\theta)$ surpasses a critical value, chosen to be 0.01.
For each quantum dot, all values of $\eta\in \{0.01, 0.05, 0.1, 0.5, 1.0\}$ were checked.
For the weakest interaction strength considered, $\lambda = 1$, setting $\eta=1$ was insufficient to cause significant circular symmetry breaking.
\label{table:eta-min}}
\end{center}
\end{table}
Electron densities for two example systems are shown in Fig.~\ref{fig:perturbation} as functions of the perturbation strength $\eta$.
In the limit as $\eta \rightarrow 0$, both have ground states with angular momentum $L=0$.
The number of blobs appearing when the symmetry is broken matches the number of particles, showing that the localization is not a trivial interference effect.

\section{Conclusion}
\label{chap:conclusion}

Most papers on Wigner molecules argue that they resemble Wigner crystals in the homogeneous electron gas and are caused by strong interactions and correlations.
As we demonstrated mathematically and computationally, however, the states identified as static Wigner molecules in the literature to date are caused by simple interference effects that have nothing to do with interactions.
They can even be seen in non-interacting systems.
We also discussed how sudden changes in the symmetry of the ground state of a quantum dot can be explained in terms of level crossings and a modified version of Hund's rules.
Level crossings are not analogous to spontaneous symmetry breaking at a continuous phase transition and do not lead to the formation of Wigner molecules.

We then gave a better definition of Wigner molecules in terms of four concepts:
\begin{enumerate}
  \item There is no sharp transition between the normal and Wigner molecule regimes.
  \item If the ground state is chosen to be a simultaneous eigenstate of energy and orbital angular momentum, the electron density must always be rotationally invariant.
    Angular modulations, often called ``rotating Wigner molecules'', may be seen in the pair-correlation function and grow smoothly in strength with the interaction.
  \item If the interaction is strong enough, or, equivalently, if the frequency $\omega$ associated with the parabolic potential is low enough, eigenstates with many different center-of-mass orbital angular momenta lie close in energy to the ground state.
Thus, applying a small symmetry-breaking perturbation is sufficient to mix states with a wide range of center-of-mass angular momenta and collapse the system into a strongly symmetry-broken state.
The non-uniform angular density that results constitutes a static Wigner molecule, caused by strong electron-electron correlations.
  \item In the limit as the interaction strength tends to infinity ($\omega \rightarrow 0$), the strength of the symmetry-breaking perturbation required to create a static Wigner molecule with strong angular modulations tends to zero.
\end{enumerate}
Concepts 2 and 4 were demonstrated mathematically in Appendices \ref{chap:so2symm} and \ref{chap:separation} and all four concepts were verified computationally.
To the best of our knowledge, true static Wigner molecules as defined above have not previously been observed.

Finally, our NNVMC simulations have provided new benchmark results for the energies, electron densities and pair-correlation functions of quantum dots with up to 20 electrons. More systems can be found in the supplementary information. The new benchmarks are more accurate than the CCSD(T) and DMC results already available in the literature.\cite{lohne2011ab}

\section{Supplementary Material}
The authors have provided three kinds of supplementary material. 
The first is a document which explains the additional data available. 
It also contains a table (Table S1) with ground state energies of several fully-polarized quantum dots. 
This table is also available as a CSV file in the supplementary material and FigShare. 
Lastly, we provide the electron density plots of the systems available in the supplementary material. 

\section{Author Contributions}
Conceptualization, APF, GC, HS and WMCF; Data curation, APF; Formal analysis, APF, GC, HS and WMCF; Investigation, APF, GC and WMCF; Methodology, APF, GC and WMCF; Software, GC and HS; Validation, APF; Visualization, APF; Supervision, WMCF; Funding Acquisition WMCF; Writing - original draft, APF; Writing - review and editing, APF, WMCF.

\section{Funding}

APF's PhD work is supported by the UK Engineering and Physical Sciences Research Council under grant EP/W524323/1.
Calculations were carried out with resources provided by the Baskerville Accelerated Compute Facility through a UK Research and Innovation Access to HPC grant.
The authors acknowledge the Gauss Centre for Supercomputing e.V.\ (www.gauss-centre.eu) for providing computing time through the John von Neumann Institute for Computing (NIC) on the GCS Supercomputer JUWELS at J\"{u}lich Supercomputing Centre (JSC), and the EUROfusion consortium for providing computing time on the Leonardo Supercomputer at CINECA in Bologna.
The EUROfusion Consortium is funded by the European Union via the Euratom Research and Training Programme
(Grant Agreement No 101052200 — EUROfusion).
Views and opinions expressed are however those of the authors only and do not necessarily reflect those of the European Union or the European Commission.
Neither the European Union nor the European Commission can be held responsible for them. 

\section{Conflicts of Interest}
The authors declare no conflict of interest. The funding sponsors had no role in the design of the study; in the collection, analyses, or interpretation of data; in the writing of the manuscript, and in the decision to publish the results.

\section{Data Availability Statement}
The original contributions presented in the study are included in the article and supplementary material, further inquiries can be directed to the corresponding authors. 

\appendix

\section{Electron density under SO(2) symmetry}
\label{chap:so2symm}

Electronic angular modulations in 2D quantum dots were first seen using DFT within the local spin density approximation (LSDA), which showed a spin-density wave.\cite{koskinen1997hund, borgh2005correlation}
Unrestricted Hartree-Fock showed both a spin-density wave and a charge-density wave.
However, this was soon explained in terms of the approximations made.\cite{ghosal2007incipient}
Due to the $SO(2)$ symmetry of the Hamiltonian under a collective rotation of all of the electron coordinates, the set of ground states must transform under an irreducible representation of $SO(2)$.
As $SO(2)$ is an Abelian group, its representations are one dimensional and can be written as $e^{iL\theta}$.
Hence
\begin{align}
  \psi(R(\theta)\bm{r}_{1}, R(\theta)\bm{r}_{2}, \dots, R(\theta)\bm{r}_{N}) = 
  e^{iL\theta}\psi(\bm{r}_{1}, \bm{r}_2, \ldots, \bm{r}_{N}),
\end{align}
where $L$, the total angular momentum, takes integer values. 
We show below that all such states produce a rotationally uniform one-electron density
\begin{align}
    n(\bm{r}) &= N\int d^{2}\bm{r}_{2}\dots d^{2}\bm{r}_{N}|\psi(\bm{r}, \bm{r}_{2}, \dots, \bm{r}_{N})|^2.
\end{align}

Consider a rotation of the electron density,
\begin{align}
    &n(R(\theta)\bm{r}) =\nonumber\\
    &=N\int d^{2}\bm{r}_{2}\dots d^{2}\bm{r}_{N}|\psi(R(\theta)\bm{r}, \bm{r}_{2}, \dots, \bm{r}_{N})|^2 \nonumber\\
    &= N\int d^{2}\bm{r}_{2}\dots d^{2}\bm{r}_{N}|\psi(R(\theta)\bm{r}, R(\theta)\bm{r}_{2}, \dots, R(\theta)\bm{r}_{N})|^2 \nonumber\\
    &= N\int d^{2}\bm{r}_{2}\dots d^{2}\bm{r}_{N}|e^{iL\theta}\psi(\bm{r}, \bm{r}_{2}, \dots, \bm{r}_{N})|^2 = n(\bm{r}),
\end{align}
where $R(\theta)$ is a 2D rotation matrix, and we have used $\det(R(\theta))=1$.
Therefore, there always exists a basis for the possibly degenerate space of ground states that has a circularly symmetric electron density.
This implies that the charge-density waves observed are artifacts of picking a particular (in most cases real-valued) superposition of the degenerate ground states.
They are not interaction-driven effects.
As was shown in Fig.~\ref{fig:pdf_real_choice}, they can even be seen in the non-interacting limit.

\section{Separation of Variables}
\label{chap:separation}

This appendix demonstrates that, at low enough $\omega$, there exist a large number of states near the ground state in energy, with different center-of-mass angular momenta.
Start with
\begin{align}
    H &= \sum_{i}\left(-\frac{\hbar^2}{2m}\nabla_{r_{i}}^{2}+\frac{1}{2}m\omega^{2}r_{i}^{2}\right) + \sum_{i<j}\frac{e^2}{4\pi\epsilon_0 r_{ij}}.
\end{align}
Consider the following coordinate transformation
\begin{align}
    \begin{pmatrix}\bm{r}_c\\\bm{\Delta}_{1}\\\bm{\Delta}_{2}\\\vdots\end{pmatrix}=\frac{1}{N}\begin{pmatrix}1&1&1&\dots\\1&-1&0&\dots\\0&1&-1&\dots\\\vdots&\vdots&\vdots&\ddots\end{pmatrix}\begin{pmatrix}\bm{r}_{1}\\\bm{r}_{2}\\\bm{r}_{3}\\\vdots\end{pmatrix}\label{eq:A},
\end{align}
where $\bm{r}_c$ is the center of mass and $\bm{\Delta}_{i} = \frac{1}{N}(\bm{r}_{i} - \bm{r}_{i+1})$ are difference coordinates.
For convenience, we define $A$ to be the matrix in Eq.~(\ref{eq:A}), including the $1/N$ prefactor.
Noting that 
\begin{align}
    AA^{T}=\frac{1}{N^{2}}\begin{pmatrix}N&0&0&0&\dots\\0&2&-1&0&\dots\\0&-1&2&-1&\dots\\0&0&-1&2&\dots\\\vdots&\vdots&\vdots&\vdots&\ddots\end{pmatrix},
\end{align}
we can write 
\begin{align}
    \frac{\hbar^2}{2m}\sum_{i=1}^{N}\nabla_{r_{i}}^{2}&=\frac{\hbar^2}{2mN}\nabla_{\bm{r}_c}^{2}+\frac{\hbar^2}{mN^{2}}\sum_{i=1}^{N-1}\nabla_{\bm{\Delta}_{i}}^{2}\nonumber\\
    &-\frac{\hbar^2}{mN^{2}}\sum_{i=1}^{N-2}\bm{\nabla}_{\bm{\Delta}_{i}}\cdot\bm{\nabla}_{\bm{\Delta}_{i+1}}.
\end{align}
Consider now the harmonic potential term
\begin{align}
    \frac{1}{2}m\omega^{2}\sum_{i=1}^{N}r_{i}^{2}&=\frac{1}{2}m\omega^{2}\bm{x}^{T}\bm{x},\nonumber\\
    &=\frac{1}{2}m\omega^{2}\bm{y}^{T}(A^{-1})^{T}(A^{-1})\bm{y},\nonumber\\
    &= \frac{1}{2}m\omega^{2}\bm{y}^{T}B\bm{y},
\end{align}
where 
$\bm{x}=\begin{pmatrix}\bm{r}_{1}&\bm{r}_{2}&\dots\end{pmatrix}^{T}$, $\bm{y}=\begin{pmatrix}\bm{r}_c&\bm{\Delta}_{1}&\dots\end{pmatrix}^{T}$ and $B = (AA^{T})^{-1}$.
As $AA^{T}$ is block-diagonal in the $\bm{r}_c$ sector, the form of $B$ is
\begin{align}
    B &= \begin{pmatrix}N&0\\0&C\end{pmatrix},
\end{align}
where $C$ is an $N-1$ by $N-1$ matrix.
Hence, 
\begin{align}
    \frac{1}{2}m\omega^{2}\sum_{i=1}^{N}r_{i}^{2} &= \frac{1}{2}mN\omega^{2}r_c^{2} + f(\bm{\Delta}_{1}, \dots, \bm{\Delta}_{N-1}),
\end{align}
for some complicated function $f$.
Therefore, the Hamiltonian in terms of the new coordinates is given by
\begin{align}
  H =
  &-\frac{\hbar^2}{2mN}\nabla_{\bm{r}_c}^{2} + \frac{\hbar^2}{2m}N\omega^{2}r_c^{2} 
  - \frac{\hbar^2}{mN^{2}}\sum_{i=1}^{N-1}\nabla_{\bm{\Delta}_{i}}^{2} \nonumber \\
    &+\frac{\hbar^2}{mN^{2}}\sum_{i=1}^{N-2}\bm{\nabla}_{\bm{\Delta}_{i}}\cdot\bm{\nabla}_{\bm{\Delta}_{i+1}}
    + \widetilde{f}(\bm{\Delta}_{1}, \dots, \bm{\Delta}_{N-1}),
\end{align}
where we have noted that the Coulomb term is independent of $\bm{r}_c$ and included it in $\widetilde{f}$.
The Hamiltonian therefore separates into terms that depend only on the center of mass and terms that depend only on the relative coordinates.
The Schrodinger equation for the terms involving the center of mass is 
\begin{align}
    \left(-\frac{\hbar^2}{2mN}\nabla_{\bm{r}_c}^{2} + \frac{1}{2}mN\omega^{2}r_c^{2}\right)\psi(\bm{r}_c) &= E_{c}\psi(\bm{r}_c).
\end{align}

Now perform a change of variables, introducing the dimensionless variable $\bm{q}=\sqrt{\frac{mN\omega}{\hbar}}\,\bm{r}_c$, to obtain the equation
\begin{align}
    \left(-\frac{1}{2}\nabla^{2}_{\bm{q}}+\frac{1}{2}q^{2}\right)\psi(\bm{q})=\frac{E_{c}}{\hbar\omega}\psi(\bm{q}),
\end{align}
which is solved by the Fock-Darwin states.\cite{kouwenhoven2001few}
The corresponding center-of-mass energy eigenvalues are
\begin{align}
    \frac{E_{c}}{\hbar\omega} &= 2n_{c} + |l_{c}| + 1,
\end{align}
where $l_{c} \in \mathbb{Z}$ is the center-of-mass angular momentum and $n_{c} \in \mathbb{Z}_{\geq 0}$ is the center-of-mass principal quantum number.
We are only concerned with how the energy depends on $l_c$, so set $n_{c}=0$ and ignore the zero-point energy to obtain
\begin{align}
    E_{c} &\sim \hbar\omega |l_{c}|.
\end{align}
This implies that, as $\omega\rightarrow 0$, the difference in energy between eigenstates with different values of $l_{c}$ tends to zero.
Although the part of the Schrodinger equation that depends on $\bm{\Delta}_{i}$ also contributes to the total angular momentum $L$, in the small $\omega$ limit, there exist almost gapless excitations to different $l_{c}$ and hence different $L$.
This proves that, for small enough $\omega$, the second requirement for a Wigner molecule is satisfied.

As a final comment, we have not yet considered spin.
The Hamiltonian is spin independent, so spin only comes into play in the form of exchange statistics.
As $\bm{r}_c=\frac{1}{N}\sum_{i}\bm{r}_{i}$, exchanging the positions of any two electrons has no effect on the center-of-mass wave function.
It is the $\bm{\Delta}_{i}$ part of the wave function that must be totally antisymmetric.
Hence, in the small $\omega$ limit, there are low-energy excitations corresponding to different $l_{c}$, and hence different $L$, even when spin is considered.

\newpage
\bibliography{dots}

\clearpage

\section*{Supplementary Material}

In the main text, we support our theory for the formation of Wigner molecules in quantum dots with neural-network variational Monte Carlo calculations.
As discussed, the results obtained improve upon the existing calculations found in the literature.
As such, we provide an extensive set of calculations for several quantum dots.
We use the following Hamiltonian,
\begin{align}
  \hat{\mathcal{H}} &= \sum_{i=1}^{N}\left(-\frac{1}{2}\nabla_{i}^{2}+\frac{1}{2}r_{i}^{2}\right) + \sum_{i=1}^{N}\sum_{j>i}^{N}\frac{\lambda}{|\bm{r}_{i} - \bm{r}_{j}|},
\end{align}
where
\begin{align}
  l_{0} &= \sqrt{\frac{\hbar}{m\omega}}, & 
    \lambda &= \frac{e^{2}}{4\pi\epsilon_{0}l_{0}\hbar\omega},
\end{align}
energies are measured in units of $\hbar\omega$, lengths in units of $l_{0}$, $\omega$ is the harmonic potential strength, $m$ is the electron mass, $e$ is the electron charge, $\hbar$ is the reduced Plank constant, and $\epsilon_{0}$ is the permittivity of free space.
We simulate fully spin-polarized quantum dots with a large range of electron numbers $N$.
The values of $\lambda$ are chosen such that they are logarithmically spaced from 1 to 10.
Some simulations failed to converge, these cases have been omitted.
The results can be found in Table \ref{table:data}, and in CSV format as further supplementary information.
Note that in the main text, the energies obtained were converted to Hartrees for direct comparison with the literature, but we have not done so here.
We have also calculated the single-particle densities, which can be found online as further supplementary information.\\

\begin{longtable}{>{\centering\arraybackslash}p{1cm} >{\centering\arraybackslash}p{1cm} >{\centering\arraybackslash}p{2cm}}
\caption{Ground state energies $E$ of quantum dots. $N$ is the number of particles and $\lambda$ is the Coulomb interaction strength.\label{table:data}} \\  

\toprule
$N$ & $\log_{10}(\lambda)$ & $E$ \\  
\midrule
\endfirsthead

\caption[]{(continued)} \\  
\toprule
$N$ & $\log_{10}(\lambda)$ & $E$ \\
\midrule
\endhead

\bottomrule
\endfoot

\bottomrule
\endlastfoot

\begin{filecontents*}{Data/supplementary.csv}
N,log10lambda,E
2,0.0,3.062474(9)
2,0.25,3.195002(9)
2,0.5,3.596626(8)
2,0.75,4.725604(9)
2,1.0,7.52905(2)
3,0.0,5.18665(1)
3,0.25,5.58173(1)
3,0.5,6.75805(1)
3,0.75,9.95997(2)
3,1.0,17.58717(2)
4,0.0,8.35402(2)
4,0.25,9.10084(2)
4,0.5,11.32063(3)
4,0.75,17.32070(5)
4,1.0,31.5525(1)
5,0.0,11.57910(2)
5,0.25,12.79801(2)
5,0.5,16.39926(3)
5,0.75,26.04026(5)
5,1.0,48.769(8)
6,0.0,14.86210(2)
6,0.25,16.67181(2)
6,0.5,21.98945(4)
6,0.75,36.08534(6)
6,1.0,69.0283(2)
7,0.0,19.16645(2)
7,0.25,21.61424(2)
7,0.5,28.79437(5)
7,0.75,47.78443(8)
8,0.0,23.52427(3)
8,0.25,26.71886(3)
8,0.5,36.06438(5)
8,0.75,60.68880(8)
8,1.0,118.13(3)
11,0.0,37.87611(3)
11,0.25,43.88140(3)
11,0.5,61.30126(5)
11,0.75,106.6226(1)
11,1.0,212.583(8)
12,0.0,43.39484(3)
12,0.25,50.47874(4)
12,0.5,70.99713(5)
12,0.75,124.2827(1)
13,0.0,48.97396(3)
13,0.25,57.25686(4)
13,0.5,81.19072(6)
13,0.75,143.1121(1)
18,0.0,80.39300(4)
18,0.25,95.74364(4)
18,0.5,139.74746(8)
19,0.0,87.19996(4)
19,0.25,104.21119(4)
19,0.5,152.89083(6)
20,0.0,94.05612(3)
20,0.25,112.82784(8)
20,0.5,166.43775(6)
27,0.0,148.86893(5)
27,0.25,181.62817(6)
27,0.5,274.4751(1)
\end{filecontents*}
\csvreader[
    head to column names,
    late after line=\\
]{Data/supplementary.csv}{N=\csvcoli, log10lambda=\csvcolii, E=\csvcoliii}{%
    \csvcoli & \csvcolii & \csvcoliii
}

\end{longtable}


\end{document}